\documentclass[twocolumn,3p,preprint]{elsarticle}
\journal{Physics Letters B}
\usepackage[numbers]{natbib}
\RequirePackage{fancyhdr}
\RequirePackage{graphicx}
\usepackage{placeins}
\RequirePackage{adjustbox}
\usepackage{rotating}
\usepackage{multirow}
\usepackage{geometry}
\usepackage{ragged2e}
\usepackage{color}
\usepackage{array}
\usepackage{float}
\RequirePackage{amssymb}
\RequirePackage{epstopdf}
\RequirePackage{amsmath, amsfonts}
\biboptions{sort&compress}
\bibpunct{[}{]}{,}{n}{,}{,}

\begin{document}
\begin{frontmatter}

\title{Deformation dependence of the isovector giant dipole resonance:\\ The neodymium isotopic chain revisited}

\author[Wits,iTL]{L.M.~Donaldson}\ead{lindsay.donaldson18@gmail.com}
\author[TAM]{C.A. Bertulani}
\author[Wits]{J.~Carter}
\author[Dubna]{V.O.~Nesterenko}
\author[TUDarm]{P.~von~Neumann-Cosel\corref{cor1}}\ead{vnc@ikp.tu-darmstadt.de}
\author[iTL]{R.~Neveling}
\author[TUDarm]{V.Yu.~Ponomarev}
\author[Erlangen]{P.-G.~Reinhard}
\author[Wits]{I.T.~Usman}
\author[iTL,US]{P.~Adsley}
\author[US]{J.W.~Brummer}
\author[iTL]{E.Z.~Buthelezi}
%\author[iTL]{J.L. Conradie}
\author[WitsGeo]{G.R.J.~Cooper}
\author[UCT]{R.W.~Fearick}
\author[iTL]{S.V.~F\"ortsch}
%\author[iTL]{D.T.~Fourie}
\author[RCNP]{H.~Fujita}
\author[OsakaU]{Y.~Fujita}
\author[Wits]{M.~Jingo}
\author[Dubna]{W.~Kleinig}
\author[Wits]{C.O.~Kureba}
\author[Prague]{J.~Kvasil}
\author[Wits]{M.~Latif}
\author[US]{K.C.W.~Li}
\author[iTL]{J.P.~Mira}
\author[iTL]{F.~Nemulodi}
\author[iTL,US]{P.~Papka}
\author[Wits,iTL]{L.~Pellegri}
\author[TUDarm]{N.~Pietralla}
\author[TUDarm]{A.~Richter}
\author[Wits]{E.~Sideras-Haddad}
\author[iTL]{F.D.~Smit}
\author[iTL]{G.F.~Steyn}
\author[US]{J.A.~Swartz}
\author[RCNP]{A.~Tamii}

\address[Wits]{School of Physics, University of the Witwatersrand, Johannesburg 2050, South Africa}
\address[iTL]{iThemba LABS, P. O. Box 722, Somerset West 7129, South Africa}
\address[TAM]{Department of Physics and Astronomy, Texas A\&M University-Commerce, Commerce, Texas 75429, USA}
\address[Dubna]{Bogoliubov Laboratory of Theoretical Physics, Joint Institute for Nuclear Research, Dubna, Moscow region, 141980, Russia}
\address[TUDarm]{Institut f\"ur Kernphysik, Technische Universit\"at Darmstadt, D-64289 Darmstadt, Germany}
\address[Erlangen]{Institut f\"ur Theoretische Physik II, Universit\"at Erlangen, D-91058 Erlangen, Germany}
\address[US]{Department of Physics, University of Stellenbosch, Matieland 7602, South Africa}
\address[WitsGeo]{School of Geosciences, University of the Witwatersrand, Johannesburg 2050, South Africa}
\address[UCT]{Department of Physics, University of Cape Town, Rondebosch 7700, South Africa}
\address[RCNP]{Research Center for Nuclear Physics, Osaka University, Ibaraki, Osaka 567-0047, Japan}
\address[OsakaU]{Department of Physics, Osaka University, Toyonaka, Osaka 560-0043, Japan}
\address[Prague]{Institute of Particle and Nuclear Physics, Charles University, CZ-18000, Prague 8, Czech Republic}
\cortext[cor1]{Corresponding author}

\begin{abstract}

Proton inelastic scattering experiments at energy $E_{\mathrm{p}} = 200$\,MeV and a spectrometer scattering angle of 0$^{\circ}$ were performed on $^{144,146,148,150}$Nd and $^{152}$Sm exciting the IsoVector Giant Dipole Resonance (IVGDR). 
Comparison with results from photo-absorption experiments reveals a shift of resonance maxima towards higher energies for vibrational and transitional nuclei. 
%The $^{150}$Nd and $^{152}$Sm nuclei are predicted to lie near the critical point of a shape phase transition with a soft quadrupole deformation potential. 
The extracted photo-absorption cross sections in the most deformed nuclei, $^{150}$Nd and $^{152}$Sm, exhibit a pronounced asymmetry rather than a distinct double-hump structure expected as a signature of $K$-splitting.
This behaviour may be related to the proximity of these nuclei to the critical point of the phase shape transition from vibrators to rotors with a soft quadrupole deformation potential.   
Self-consistent random-phase approximation (RPA) calculations using the SLy6 Skyrme force provide a relevant description of the IVGDR shapes deduced from the present data.  
 
\end{abstract}


\end{frontmatter}

\section{Introduction}
\label{sec:intro}

Giant resonances represent a prime example of collective modes in the nucleus. 
A smooth mass-number dependence of the resonance parameters is characteristic of all nuclear giant resonances and, as such, a study of them yields information about the non-equilibrium dynamics and the bulk properties of the nucleus \cite{Har01}.
The oldest and best known giant resonance is the IVGDR owing to the high selectivity for isovector E1 excitation in photo-absorption experiments. 
The properties of the IVGDR have been studied extensively using ($\gamma$,xn)-type experiments, particularly in the Saclay \cite{Ber77} and Livermore \cite{Ber75} laboratories. 
These sets of experiments are a major source of information with respect to the $\gamma$-strength function \cite{Bar72} above the neutron threshold - an important quantity used in statistical reaction calculations relevant to applications like astrophysical large-scale reaction networks \cite{Arn07,Kap11}, reactor design \cite{Cha11}, and even nuclear waste transmutation \cite{Bow98}.

Recently, a new experimental technique for the extraction of electric dipole-strength distributions in nuclei via relativistic Coulomb excitation has been developed~\cite{Tam09,Nev11}. 
It utilises proton inelastic scattering with energies of a few hundred MeV at scattering angles close to $0^\circ$. 
Although many of these experiments focus on establishing the strength below and around neutron threshold and its contribution to the dipole polarisability \cite{Tam11,Pol12,Kru15,Has15,Bir17}, such data also provide information on the photo-absorption cross sections in the energy region of the IVGDR. 
 
The chain of stable even-even neodymium isotopes is known to comprise a transition from spherical  to deformed ground states for heavier isotopes and thus represents an excellent test case to study the influence of deformation on the properties of the IVGDR. 
A ($\gamma$,xn) experiment at Saclay \cite{Car71} revealed that the width increases with deformation evolving into a pronounced double-hump structure in the most deformed nuclide $^{150}$Nd, considered to be a textbook example \cite{Boh75} of $K$-splitting owing to oscillations along the different axes of the quadrupole-deformed ground state. 
Here, we report new photo-absorption cross sections for $^{144,146,148,150}$Nd extracted from 200 MeV proton scattering experiments with results differing significantly from Ref.~\cite{Car71}. 
In particular, no double-hump structure is observed in the most deformed $^{150}$Nd nucleus.
This finding is confirmed in a further measurement of the comparably deformed $^{152}$Sm nucleus, again in contrast  to a ($\gamma$,xn) measurement at Saclay \cite{Car74}. 
%Together with the conclusions of a critical reanalysis of the experimental methods applied in the Saclay experiments \cite{Var14} and new ($\gamma$,n) data \cite{Nyh15, Fil14} the results presented raise general questions regarding the ($\gamma$,xn) results obtained at Saclay and their reliability for applications.
This unexpected result may be related to the special structure of these two nuclei which are predicted to lie near the critical point \cite{Cas01} of a shape phase transition from spherical to quadrupole-deformed ground states \cite{Iac01}.

\section{Experiment and analysis}
\label{sec:exp-results}

The proton inelastic scattering experiments were performed at the iThemba Laboratory for Accelerator Based Sciences (iThemba LABS) in South Africa. 
The K600 magnetic spectrometer, positioned at 0$^{\circ}$ with the acceptance defined by a circular collimator having an opening angle \mbox{$\theta_{\rm lab} = \pm$\,1.91$^{\circ}$}, was used to analyse a scatterered 200\;MeV dispersion-matched proton beam delivered from the Separated Sector Cyclotron of iThemba LABS. 
The self-supporting $^{144,146,148,150}$Nd and $^{152}$Sm targets were all isotopically enriched to values $>$96\% (except $^{148}$Nd enriched to 90\%) with areal densities ranging from 1.8 to 2.6~mg/cm$^2$. 
The corresponding ground-state deformation parameters $\beta_2$ are given in the second column of Table \ref{table:GammaParam}.
The beam preparation and the detector setup are described in Ref.~\cite{Nev11}. 
Details regarding the data extraction and analysis of the present measurements can be found in Ref.~\cite{Don16}. 

\begin{figure}[t]
\begin{center}
	\includegraphics[width=\columnwidth]{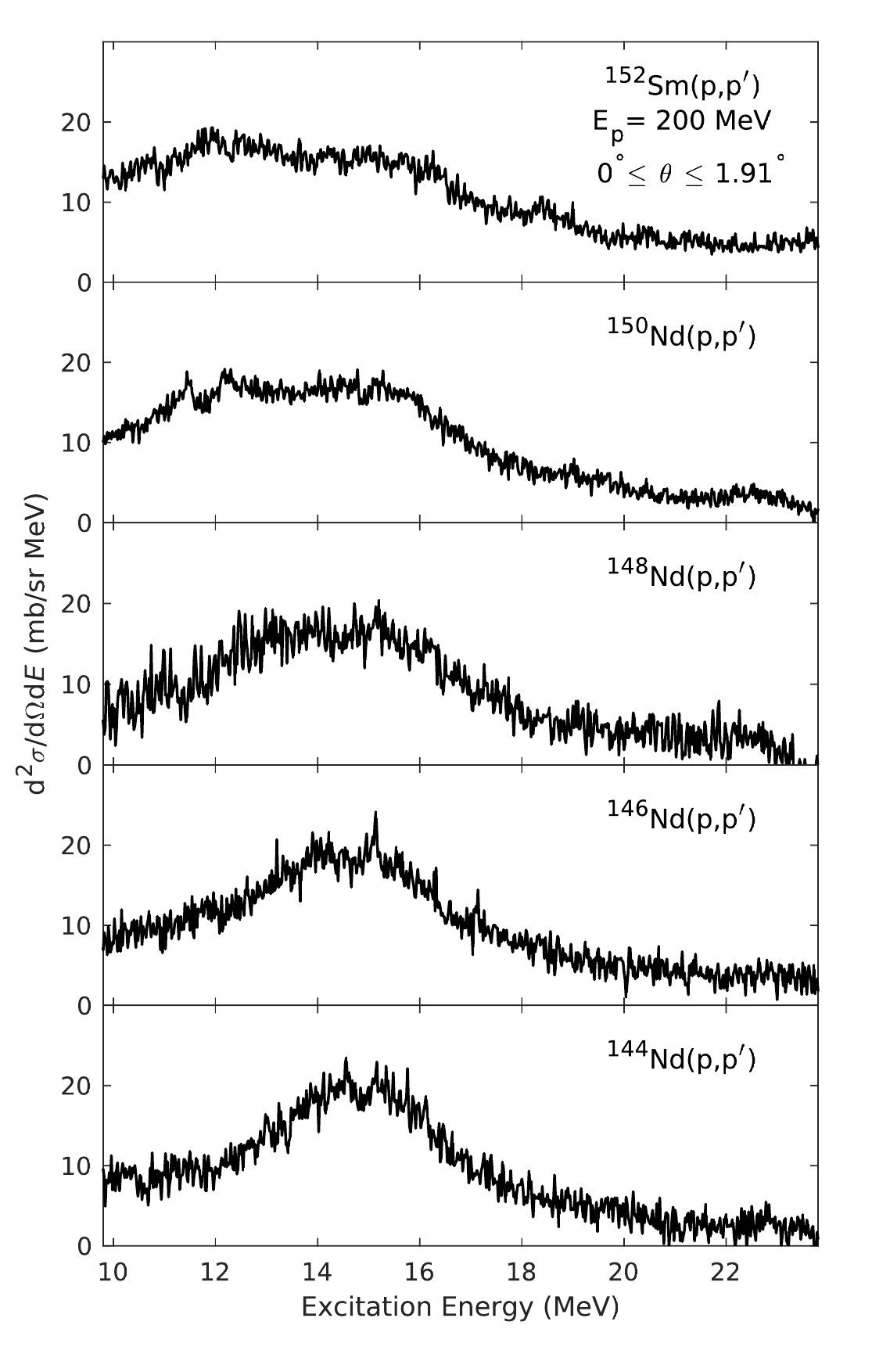}
	\caption{Experimental double-differential cross sections for the $^{144, 146, 148, 150}$Nd(p,p$^\prime$) and $^{152}$Sm(p,p$^{\prime}$) reactions at $E_{\mathrm{p}}=200$\;MeV and $\theta_{\mathrm{lab}}=0^{\circ} \pm 1.91^{\circ}$.}
	\label{fig:DDCSAll-20keV}
\end{center}
\end{figure}

In the chosen kinematic conditions, relativistic Coulomb excitation of the target nuclei is the dominant reaction mechanism. 
The resulting double-differential cross sections (with a systematic uncertainty of $\pm 7$\%) obtained following the procedures detailed in Ref.~\cite{Don16} are displayed in Fig.~\ref{fig:DDCSAll-20keV} for 20 keV energy bins.
A typical energy resolution $\Delta E = 45$ keV (FWHM) was achieved.
The broad structure visible in all spectra between approximately $E_{\rm x} = 12$ and 18\;MeV corresponds to the excitation of the IVGDR. 
Statistical errors in this region are of the order of 2-4\%.
%The sharp peak at $E_{\rm x} \approx 15$ MeV visible in some spectra results from the M1 transition in $^{12}$C to a state at $E_{\rm x} =15.11$ MeV \cite{ajz90} prominently excited in the (p,p$^\prime$) reaction at small momentum transfers. 
From  Fig.~\ref{fig:DDCSAll-20keV} it is immediately evident that the width of the IVGDR increases steadily from the nearly spherical $^{144}$Nd nucleus through the transition region to the more deformed $^{150}$Nd and $^{152}$Sm nuclei. 

\begin{figure}[tbh]
\centering
  \includegraphics[width=\columnwidth]{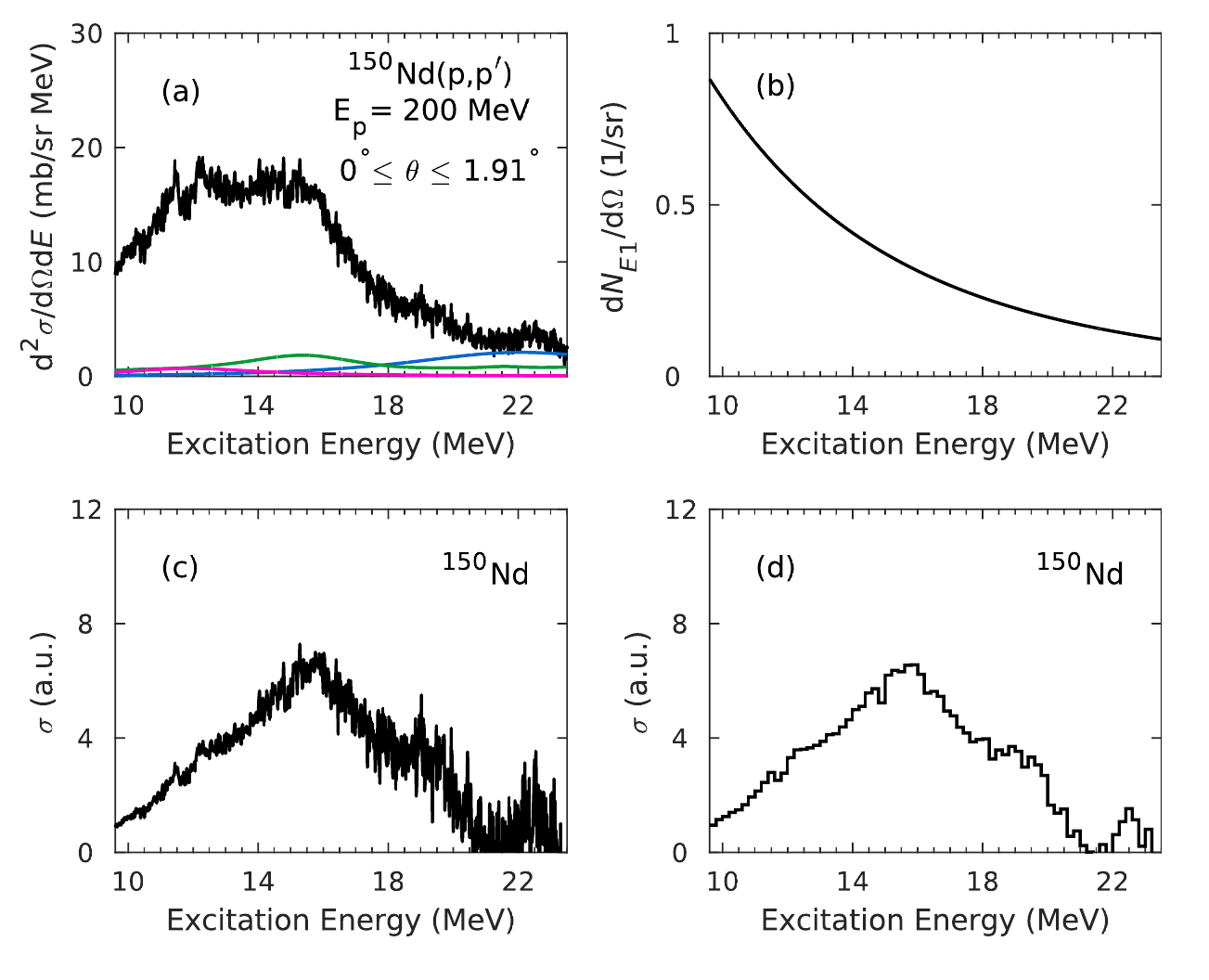}
  \caption{(Colour online). Overview of the conversion process from $^{150}$Nd(p,p$^{\prime})$ to photo-absorption cross sections: (a) double-differential (p,p$^{\prime}$) cross section and background components. The green and pink lines describe the contribution from the ISGMR and ISGQR, respectively, and the blue line the phenomenological component explained in the text; (b) virtual photon spectrum; (c) equivalent photo-absorption spectrum resulting from Eq.~(\ref{eq:Lorentz}); (d) equivalent photo-absorption spectrum rebinned to 200 keV for comparison with the photo-absorption data of Ref.~\cite{Car71}.}
  \label{fig:150NdConversion}
\end{figure}

In order to compare to the ($\gamma$,xn) data of Carlos et al.~\cite{Car71,Car74}, the (p,p$^{\prime}$) spectra had to be converted to equivalent photo-absorption cross sections. 
By way of example, Fig.~\ref{fig:150NdConversion} provides an overview of the conversion process for $^{150}$Nd. 
It can be divided into three distinct stages, namely, background subtraction in the region of the IVGDR, calculation of the virtual-photon spectrum and the division by this spectrum multiplied through by the virtual $\gamma$ energy to obtain equivalent photo-absorption cross sections.  
This procedure has been tested for several cases ($^{48}$Ca, $^{120}$Sn, $^{208}$Pb) and fair agreement of the resulting shape and absolute values of the photo-absoprtion cross sections with experiments using real photons was obtained \cite{Tam11,Pol12,Kru15,Has15,Bir17}. 

\begin{figure}[tbh]
\begin{center}
	\includegraphics[width=6cm]{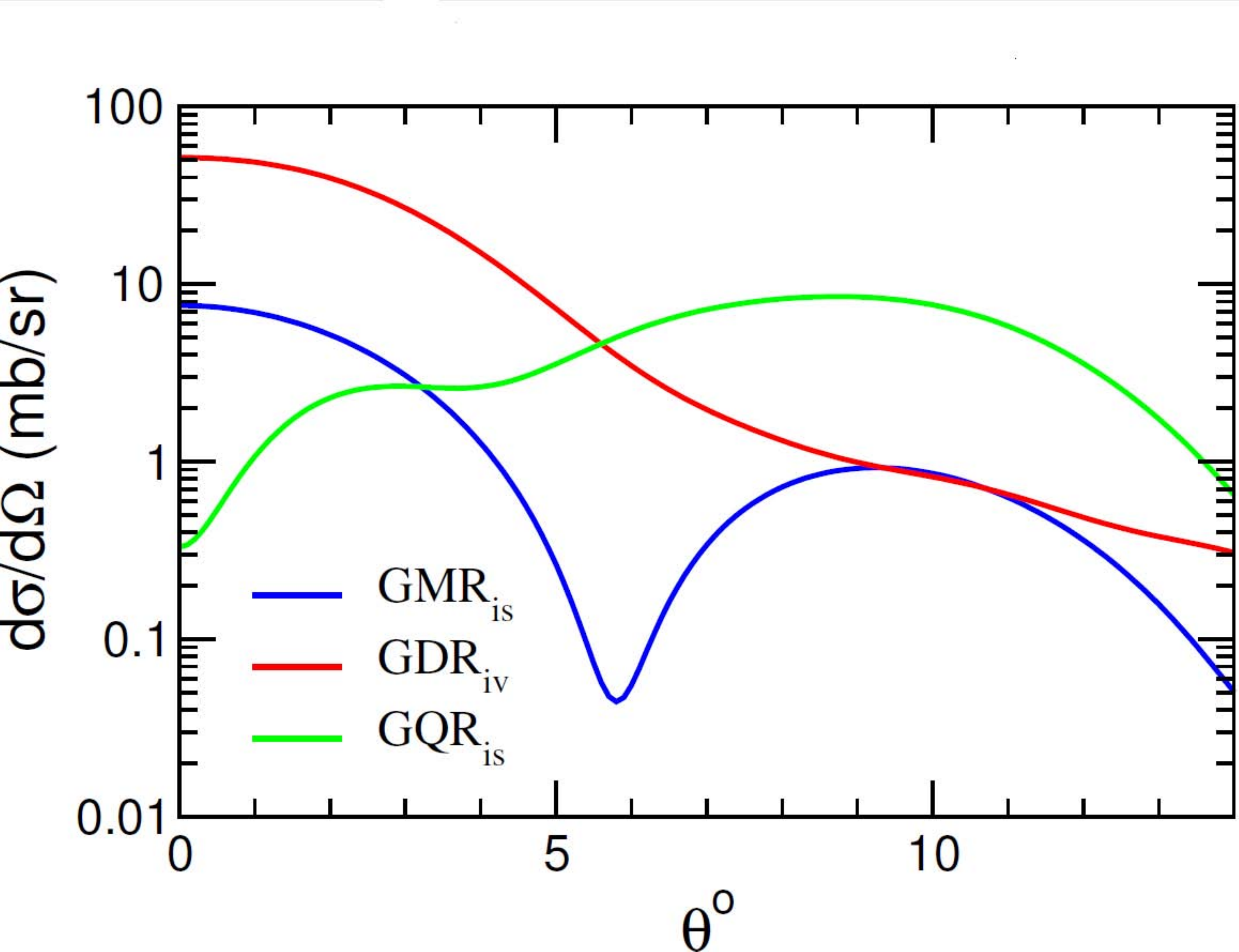}
	\caption{(Colour online). Example of DWBA calculations of the IVGDR (red), ISGMR (blue) and ISGQR (green) excitation cross sections in (p,p$^\prime)$ scattering at $E_0 = 200$ MeV off Nd and Sm isotopes.}
	\label{fig:DWBA}
\end{center}
\end{figure}

Background from nuclear processes studied in similar experiments at 300\;MeV has been found to be small in heavy nuclei \cite{Tam11,Pol12,Kru15,Has15}. 
It was modelled in the present case by three components. 
The contributions of the IsoScalar Giant Monopole Resonance (ISGMR, green line) and IsoScalar Giant Quadrupole Resonance (ISGQR, pink line) to the spectrum of Fig.~\ref{fig:150NdConversion}(a)  were estimated in the following way \cite{Kru15,Bir17}:
Theoretical angular distributions of the ISGMR and ISGQR cross sections were determined by distorted wave Born approximation calculations with the code DWBA07 \cite{dwba07} using quasiparticle phonon model (QPM) transition amplitudes and the Love-Franey effective interaction \cite{Fra81} as input (analogous to Ref.~\cite{Pol12}).
A representative example of such calculations for Nd and Sm isotopes is shown in Fig.~\ref{fig:DWBA}.

After averaging over the experimental angular acceptance, these calculations provide a relation between theoretical  cross sections and transition strengths under the assumption of a dominant one-step reaction mechanism, which should be well fulfilled at an incident proton energy of 200 MeV.
Utilising this proportionality, experimental ISGMR and ISGQR strength distributions can then be converted to (p,p$^\prime$) cross sections in the present spectra.
For comparison, the predicted IVGDR cross section is also shown and clearly dominates the spectra.
However, the B(E1) transition strengths (and thus the photo-absorption cross sections) cannot be extracted with this method because the Coulomb-nuclear interference term breaks the proportionality. 
  
Itoh et al.~\cite{Ito03} reported isoscalar giant resonance strength distributions for the Sm isotope chain which could be directly applied to $^{152}$Sm.
The results of Ref.~\cite{Ito03} were also used for the corresponding Nd isotones, which show very similar deformation parameters, with a correction for the global mass dependences of the ISGMR and the ISGQR \cite{Har01}.     
The ISGQR contribution was independently estimated from a recent study of the ISGQR in the Nd isotope chain \cite{Kur17} with methods analogous to Refs.~\cite{She04,She09,Usm11} in good correspondence with results from the above procedure. 

A phenomenological background shown as blue line Fig.~\ref{fig:150NdConversion}(a) describes the behaviour of the double-differential cross section at the high excitation energy part of the spectrum where the Coulomb excitation contribution is negligible. 
This component incorporates all unknown multipolarity contributions as well as quasi-free scattering and is approximated by finding the maximum of the cross section between $E_{\rm x} = 20$ MeV and 23 MeV and using a width that best describes the spectrum in this region.
 A similar description for the shape of this component was found in a study of $^{208}$Pb \cite{Pol14} where an experimental extraction of the angular distribution of the background was possible. 

The virtual E1 photon spectrum \cite{Ber88} for each isotope was calculated using the eikonal approximation~\cite{Ber93} and averaged over the angular acceptance of the detector. 
The equivalent photo-absorption spectrum (cf.\ Fig.~\ref{fig:150NdConversion}(c)) was then obtained  using the following equation
\begin{equation}
\frac{d^2{\sigma}}{d{\Omega}dE_{\gamma}}={\frac{1}{E_{\gamma}}} {\frac{dN_{E1}}{d{\Omega}}} \sigma^{\pi \lambda}_{\gamma} (E_{\gamma}).
\label{eq:Lorentz}
\end{equation}
Finally,  Fig.~\ref{fig:150NdConversion}(d) shows the equivalent photo-absorption spectrum rebinned to 200 keV for direct comparison with the ($\gamma$,xn) results.
The present setup at $\theta_{\rm lab} = 0^\circ$ does not allow for the determination of accurate vertical scattering angles, thus limiting the angular resolution \cite{Nev11}.
We therefore refrain from extracting absolute photo-absorption cross sections.
The excitation energy dependence of the conversion, however, is not affected.

\section{Comparison with ($\gamma$,xn) results}
\label{sec:gamcomp}

\begin{figure}[tbh]
\begin{center}
	\includegraphics[width=5cm]{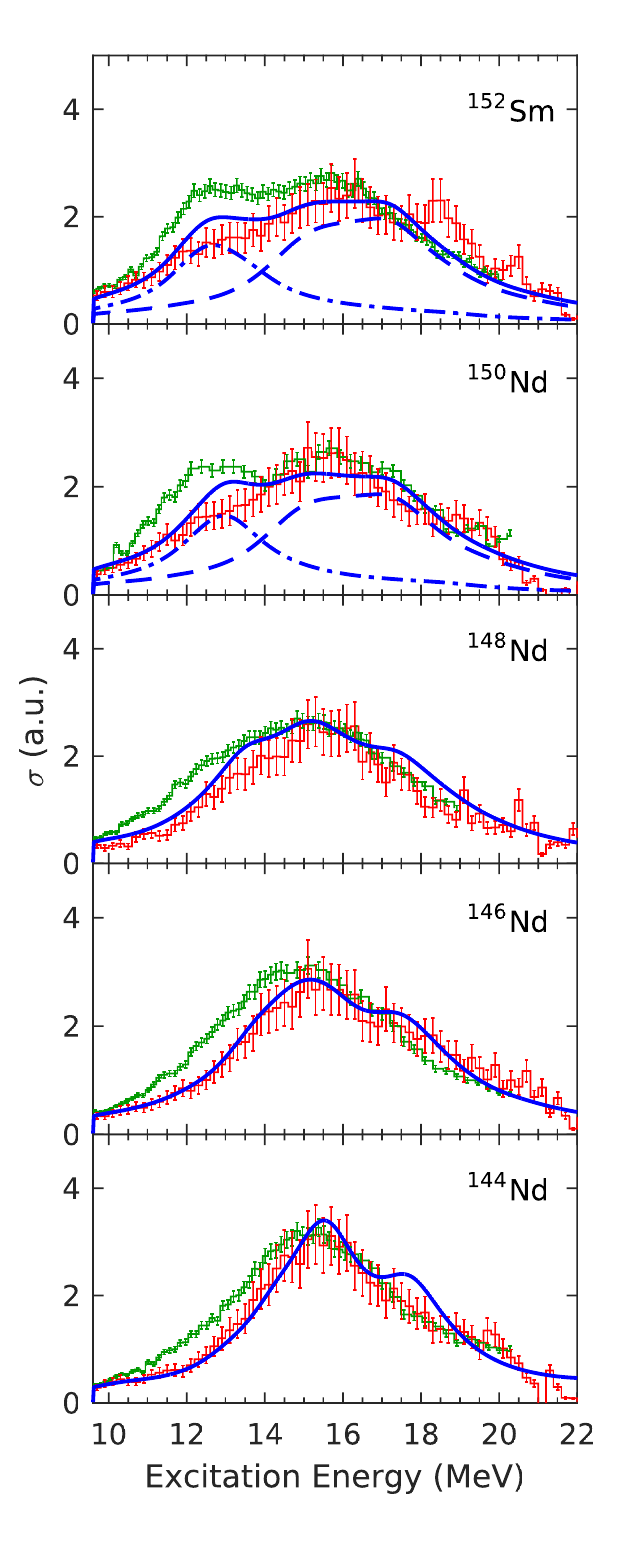}
	\caption{(Colour online). Photo-absoprtion cross sections from the present data (red histograms) normalised to the maximum of the pre-existing $(\gamma$,xn) results (green histograms) \cite{Car71,Car74}. 
The blue lines show  the results of SSRPA calculations with the SLy6 force described in the text (solid: full, dashed-dotted: $K = 0$ part, dashed: $K = 1$ part.}
	\label{fig:CvsLMD-Norm}
\end{center}
\end{figure}

Through the simultaneous measurement of the partial photo-nuclear cross sections $\sigma(\gamma,\mathrm{n})+\sigma(\gamma,\mathrm{pn})$ and $\sigma(\gamma,2\mathrm{n})$ using a monochromatic photon beam, total photo-absorption cross sections can be determined in heavy nuclei.
Data obtained with this method for the IVGDR in the stable even-even neodymium and samarium isotopic chains are given in Refs.~\cite{Car71} and \cite{Car74}, respectively. 
Figure~\ref{fig:CvsLMD-Norm} displays the rebinned spectra from the present work normalised to the maximum of the photo-absorption cross sections~\cite{Car71,Car74} to facilitate a comparison of the evolution of the shape of the IVGDR with increasing deformation.

Carlos et al.~\cite{Car71,Car74} (green histograms) observed a spreading of the IVGDR as the nuclei become softer followed by a splitting of the IVGDR into two distinct dipole modes for $^{150}$Nd and $^{152}$Sm, which were interpreted as $K=0$ and $K=1$ components.
The equivalent photo-absorption cross sections from the present work (red histograms) display a similar trend, i.e., a general broadening of the IVGDR with increasing deformation.
For the most deformed $^{150}$Nd and $^{152}$Sm, the resonance becomes skewed with increased strength on the low-energy side, but no split into two distinct components is observed. 

\begin{table*}[tbh]
\setlength{\extrarowheight}{2.5pt}
\centering
\caption{Comparison of Lorentzian parameterisations, Eq.~(\ref{eq:singleLor}), for the present photo-absorption cross sections with those from Ref.~\cite{Car71} for the neodymium isotopes and from Ref.~\cite{Car74} for $^{152}$Sm.
The quantity $R$ denotes the ratio of photo-absorption cross sections summed in the excitation energy regions $10-14$ MeV and  $14-18$ MeV, respectively.}
\label{table:GammaParam}
%\begin{adjustbox}{width=0.95\textwidth}{
\begin{tabular}{cccccccc}
\\
\hline 
Isotope & $\beta_2$ \cite{Ram01} &  $\mathit{E}_1$ (MeV) & ${\Gamma}_1$ (MeV) & $\mathit{E}_2$ (MeV) & ${\Gamma}_2$ (MeV) & $R$ & Reference \\ \hline 
%\multirow{2}{*}{$^{142}$Nd} & \multirow{2}{*}{0.09} &14.95$\pm$0.10 & 359$\pm$15 & 4.43$\pm$0.20 &  &  &  & \cite{Car71}\\
% & 14.94 & 358.3 & 4.41 &  &  &  &  & \cite{RIPL3}\\
% & & 15.66$\pm$0.02 &  & 4.22$\pm$0.06 &  &  &  & Present \\ \hline
\multirow{2}{*}{$^{144}$Nd} & \multirow{2}{*}{0.13} & 15.05$\pm$0.10 & 5.30$\pm$0.25 &  &  & 0.55 &\cite{Car71} \\
% & 15.04 & 315.5 & 5.25 &  &  &  &  & \cite{RIPL3} \\
 & & 15.64$\pm$0.01 & 4.93$\pm$0.03 &  &  &  0.42 & Present \\ \hline
\multirow{2}{*}{$^{146}$Nd} &\multirow{2}{*}{0.15} & 14.80$\pm$0.10 &  6.00$\pm$0.30 &  &  & 0.66 & \cite{Car71} \\
% & 14.72 & 308.9 & 5.75 &  &  &  &  & \cite{RIPL3}\\
 & & 15.69$\pm$0.02 & 6.11$\pm$0.07 &  &  &  0.47 & Present \\ \hline
\multirow{2}{*}{$^{148}$Nd} &\multirow{2}{*}{0.20} & 14.70$\pm$0.15 & 7.20$\pm$0.30 &  &  & 0.74 & \cite{Car71} \\
% & 12.78 & 110.3 & 4.03 & 15.49 & 216.0 & 5.21 & 0.395 & \cite{RIPL3} \\ 
& & 15.52$\pm$0.01 & 5.84$\pm$0.04 & & & 0.53 & Present \\ \hline
% & 12.78$\pm$0.13 & 58.5$\pm$10.0 & 2.27$\pm$0.48 & 15.31$\pm$0.08 & 343.0$\pm$8.6 & 5.19$\pm$0.12 & 0.07 & Present \\ \hline
\multirow{2}{*}{$^{150}$Nd} & \multirow{2}{*}{0.28} & 12.30$\pm$0.15 & 3.30$\pm$0.10 & 16.00$\pm$0.15 &  5.20$\pm$0.15 & 0.77 & \citep{Car71} \\
% & 12.30 & 175.4 & 3.38 & 16.03 & 220.5 & 5.12 & 0.525 & \cite{RIPL3} \\
& & 11.97$\pm$0.10 & 2.91$\pm$0.40 & 15.67$\pm$0.04 & 5.64$\pm$0.09 & 0.60 & Present \\ \hline
\multirow{2}{*}{$^{152}$Sm} & \multirow{2}{*}{0.31} & 12.45$\pm$0.10 & 3.20$\pm$0.15 & 15.85$\pm$0.10 & 5.10$\pm$0.20 & 0.87 & \cite{Car74} \\
% & 12.39 & 176.8 & 2.99 & 15.73 & 232.0 & 5.15 & 0.442 & \cite{RIPL3} \\
 & & 12.40$\pm$0.20 & 4.73$\pm$0.65 & 16.36$\pm$0.07 & 6.36$\pm$0.14 & 0.62 & Present \\ \hline 
\end{tabular}
%\end{adjustbox}
\end{table*}

The obvious discrepancies (in both the $K=0$ and $K=1$ regions) between the photo-absorption results and the present data are reflected in a change of parameters when attempting to describe the observed resonances by Lorentzians. 
The present spectra were fitted with a modified Lorentzian of the form used in Refs.~\cite{Car71,Car74}
\begin{equation}
\label{eq:singleLor}
\sigma(E) \propto \frac{1}{1+[(E^2 - {E_{\mathrm{R}}}^2)^2 / E^2{{\Gamma}^2}]},
\end{equation}
where $E_\mathrm{R}$ corresponds to the resonance centroid energy and $\Gamma_\mathrm{R}$ to the resonance half-width. 
For the more deformed nuclei, a sum of two modified Lorentzians was used. 
The best fit to the experimental data was selected such that the value for the reduced ${\chi}^2$ was optimised. 
In the case of the $^{148}$Nd isotope, it was found that the reduced ${\chi}^2$ value was not improved through the use of a two Lorentzian fit as assumed in a reanalysis~\cite{RIPL3} of the data of Ref.~\cite{Car71}.
The results are summarised in Table~\ref{table:GammaParam}.
In order to further illustrate the systematic differences between the two results, we also provide the ratio $R$ of summed photoabsorption cross sections in the excitation energy regions $10 -14$ and $14 -18$ MeV, respectively.   
 
For spherical and transitional nuclei, a shift of the centroid to higher energies is observed for the present data. 
The new parametrisations for the deformed nuclei do not yield a ratio of $\approx 0.5$ for $K=0$ and $K=1$ oscillator strengths, respectively, as expected for prolate deformed ground states~\cite{Ber75}. 
%This is a surprising result, which could perhaps be an indication that the assumption of complete quadrupole deformation in $^{150}$Nd and $^{152}$Sm is incorrect.
One should remember, however, that  $^{150}$Nd and $^{152}$Sm lie just above the shape phase transition from vibrators to axial rotors \cite{Cas01,Cas06}.
 Although they are already well deformed, their deformation potential is soft in the $\beta$ degree of freedom 
 \cite{Iac01}. 
The corresponding shape fluctuations thus enhance the width of the resonance peaks which, in turn, may hinder a clear discrimination of the $K=0$ and 1 branches.

The present results do not provide absolute photo-absorption cross sections and thus cannot distinguish whether the main discrepancies lie in the $K = 0$ or $K = 1$ region, but the agreement at high excitation energies in Fig.~\ref{fig:CvsLMD-Norm} and Table~\ref{table:GammaParam} suggest the former. 
In any case, they clearly indicate a different ratio of $K = 0$ and $ K =1$ components compared to Refs.~\cite{Car71,Car74}. 
This finding is independent of the background in the spectra due to nuclear processes shown to be small in the energy region of the IVGDR, cf.~Fig.~\ref{fig:150NdConversion}(a).
The largest component stems from the ISGMR, whose angular distribution peaks at $0^\circ$.
Even at the maximum of the ISGMR peak, the cross section contribution does not exceed 10\%.
%Furthermore, with increasing deformation the ISGMR develops a low-energy bump due to coupling to the $K = 0$ component of the ISGQR \cite{Gar80}, whose centroid energy roughly coincides with the $K = 0$ component of the IVGDR.
%Thus, the subtraction of the ISGMR cross sections hardly influences the shape of the IVGDR in the present data.

New photo-absorption data are available from ($\gamma$,n) experiments~\cite{Nyh15,Fil14} in the excitation-energy region between the neutron threshold and $E_{\rm x} \approx 13$ MeV.
A study of the Sm isotope chain finds systematically smaller cross sections than Ref.~\cite{Car74}, corroborating the present results. 
A similar investigation of $^{143-148}$Nd finds again significantly lower cross sections than Ref.~\cite{Car71} for the lighter isotopes but fair agreement  for $^{146,148}$Nd. 
Photo-absorption cross sections of $^{154}$Sm \cite{Kru14} have been deduced in a study of the E1 strength  with forward angle proton scattering analogous to the experiments described in Refs.~\cite{Tam11,Pol12,Kru15,Has15}. 
The results do show a double-hump structure but with a clear reduction in the $K = 0$ region and a slight enhancement in the $K = 1$ region compared to the Saclay data \cite{Car74}, again leading to a reduced $K = 0/K = 1$ ratio as in the present case.
These findings are qualitatively consistent with a global reanalysis of data taken with the Saclay method \cite{Var14}, which indicates that the ($\gamma$,n) cross sections are systematically too large and the ($\gamma$,2n) cross sections too small.

One may speculate whether the obsereved differences are related to the reaction mechanism (real vs. virtual photoexcitation). 
However, photo-absorption cross sections deduced from similar (p,p$^\prime$) experiments using the virtual photon method show good correspondence with ($\gamma$,xn) data in other cases, cf. Refs.~\cite{Tam11,Has15}.  

\section{Comparison with model calculations}
\label{sec:theory}
In order to investigate the role of $K=0$ and $K=1$ components further, a comparison with RPA calculations particularly suited for modelling the IVGDR is presented. 
The calculations are performed within the Skyrme Separable Random Phase Approximation (SSRPA) approach \cite{Nes06}.
The method is fully self-consistent since  both the mean field and residual interaction are derived from the same Skyrme functional. 
The residual interaction includes all the functional contributions as well as the Coulomb direct and exchange terms. 
The self-consistent factorisation of the residual interaction crucially reduces the computational effort for deformed nuclei and maintains high accuracy of the calculations \cite{Nes06,Nes08,Kle08}.

We note that various methods can be applied to the description of IVGDR and other collective excitations in deformed nuclei, see e.g. approaches with finite-range Gogny forces \cite{Mar16,Del10} and a relativistic mean field \cite{Yao14}. 
Some of these more elaborate approaches based on the generator  coordinate method take into account nuclear triaxiality \cite{Del10} or investigate the breaking of axial symmetry by using projection techniques \cite{Yao14}. 
However, either no theoretical predictions for the present problem are available \cite{Del10,Yao14} or they show rather large deviations from experiments \cite{Mar16}.

Here, the Skyrme parameterisation SLy6 \cite{SLy6} is used which was shown to provide a good description of the IVGDR in medium-heavy, deformed nuclei \cite{Kle08}. 
The code exploits the 2D grid in cylindrical coordinates. 
The axial quadrupole deformation characterised by the parameter $\beta_2$ is determined by minimisation of the total energy, and $\beta_2$ values obtained are typically close to data \cite{Kle08} when taking into account that data, deduced from B(E2) values, embrace ground state deformation plus some quantum fluctuations not included in mean field calculations.. 
We thus adopt the experimental values (cf.\ Table \ref{table:GammaParam}) for the quadrupole deformation in $^{146,148,150}$Nd and $^{152}$Sm, respectively.
For the nearly spherical $^{144}$Nd, a negligible deformation, $\beta_2$\,=\,0.001, is used. 
Nd and Sm isotopes in the transitional region, however, show very soft deformation energy surfaces, which gives a large uncertainty to the theoretical ground-state deformations. 
We have checked triaxiality with full 3D mean-field calculations and do  not find any for the nuclei considered here, in agreement with a systematic study of nuclear shapes using the Skyrme functional SkM* \cite{Sca14}.
    
Pairing is treated with delta forces at the BCS level \cite{Ben00}. 
A large two-quasiparticle basis up to $\sim$\,100 MeV is taken into account. 
The Thomas-Reiche-Kuhn energy-weighted sum rule \cite{Rin80} for isovector E1 strength is exhausted by $98-100$\%. 
%The SSRPA strength function $S({\rm E1})(E) =\sum_i E_i {\rm B(E1)}_i \xi_D(E-E_i)$ (sum over all SSRPA states) 
%
%\begin{figure}[t]
%\begin{center}
%	\includegraphics[width=\columnwidth]{Fig4_Replacement_2017_new.pdf}
%	\caption{(Colour online). Comparison of the equivalent photo-absorption spectra from the present work and the SSRPA  predictions smoothed with a width of 2 MeV.}
%	\label{fig:BroadSSRPA}
%\end{center}
%\end{figure}
%
To facilitate a comparison between the experimental results and the model calculations, the SSRPA predictions were smoothed with a width $\Gamma = 2$ MeV, which provides a good description of the broad structure of experimental IVGDR strength distributions in many heavy deformed nuclei \cite{Kle08}.
%   
%There is excellent agreement with experiment for all of the isotopes in question reproducing the shape of the IVGDR but also the absolute photo-absorption cross sections. 
%In the case of the more deformed $^{150}$Nd and $^{152}$Sm nuclei, the lack of an exaggerated double-peaked structure is reproduced by the SSRPA predictions.

The resulting photo-absorption cross sections are shown in Fig.~\ref{fig:CvsLMD-Norm} as blue lines.
They are normalised to the data at the high-energy flank of the IVGDR, where the results of Carlos et al.~\cite{Car71,Car74} and the present work agree reasonably well.
For the most deformed nuclei, $^{150}$Nd and $^{152}$Sm, the separation into $K = 0$ (dashed-dotted) and $K = 1$ (dashed) components is additionally shown.
For the spherical and transitional nuclei, $^{144,146,148}$Nd, the calculations are in better agreement with the  present results, i.e. favouring smaller cross sections on the low-energy flank.  
For $^{150}$Nd and $^{152}$Sm, the SSRPA results display a double-hump structure, but again with a lower $K = 0$ component than observed in the Saclay results and total cross sections closer to the present data.
Since there is a certain degree of freedom in the normalisation one could bring the theoretical results in better agreement with the results of Carlos et al.\ at lower excitation energies, but at the price of overshooting all available data at higher $E_{\rm x}$. 

\section{Conclusions}
\label{sec:conclusion}

A measurement of the (p,p$^{\prime}$) reaction at $E_{\rm p} = 200$ MeV and $\theta_{\rm lab} = 0^\circ$ favouring relativistic Coulomb excitation in the energy region of the IVGDR has been presented for the even-even $^{144-150}$Nd isotopic chain as well as for $^{152}$Sm.
While the high energy-resolution data show considerable fine structure (even in the deformed isotopes), which carries information on the role of different decay mechanisms of the giant resonances \cite{She04,She09,Usm11,Pol14} and level densities \cite{Pol14,Usm11a}, the present work focuses on a study of the evolution of the IVGDR as a function of deformation.

A general broadening of the IVGDR is observed with increasing deformation and the most deformed $^{150}$Nd and $^{152}$Sm nuclei exhibit a prononounced asymmetry rather than a double-hump structure owing to $K$-splitting, in contrast to previous photo-absorption data from Saclay \cite{Car71,Car74}.
This is interpreted as a signature of the peculiar nature of these two nuclei which lie close to the critical point of a shape phase transition from vibrators to rotators characterised by a soft potential in the $\beta$ degree of freedom \cite{Iac01}.  
Self-consistent RPA model calculations with the Skyrme SLy6 force, particularly suited to describe the IVGDR, provide a fair description of the data consistent with a reduction of cross sections on the low-energy side of the resonance with respect to the Saclay data.
In view of their general relevance, an independent test of these unexpected results would be highly valuable.
It should be possible to realise such experiments in the near future at the low-enery tagger system NEPTUN at the S-DALINAC \cite{Sav10} and at ELI-NP \cite{Fil15}.

\section*{Acknowledgements}
We thank J.L.~Conradie and the accelerator team at iThemba LABS for providing excellent beams.
We are indebted to M.~Itoh for providing us with the numerical results of Ref.~\cite{Ito03}.
%V.Yu.~Ponomarev is thanked for his help with the DWBA calculations and useful discussions.
This work was supported by the South African NRF and by the DFG under contract No.\ SFB 1245.
C.A.B.\ acknowledges support by the U.S.\ DOE grant DE-FG02-08ER41533 and the U.S.\ NSF Grant No.\ 1415656 and J.K.\ by the Czech Science Foundation (Grant No.\ P203-13-07117S).

\end{document}